\documentstyle[prl,aps,preprint]{revtex}
\tighten

\newcommand{\newc}{\newcommand}
\newc{\gsim}{\lower.7ex\hbox{$\;\stackrel{\textstyle>}{\sim}\;$}}
\newc{\lsim}{\lower.7ex\hbox{$\;\stackrel{\textstyle<}{\sim}\;$}}
\newc{\ie}{{\it i.e.}}		
\newc{\etal}{{\it et al.}}
\newc{\eg}{{\it e.g.}}		
\newc{\kev}{\hbox{\rm\,keV}}		
\newc{\mev}{\hbox{\rm\,MeV}}		
\newc{\gev}{\hbox{\rm\,GeV}}		
\newc{\tev}{\hbox{\rm\,TeV}}
\newc{\xpb}{\hbox{\rm\, pb}}
\newc{\xfb}{\hbox{\rm\, fb}}

\newc{\mtop}{m_t}
\newc{\mbot}{m_b}
\newc{\mz}{m_Z}
\newc{\mw}{m_W}
\newc{\alphasmz}{\alpha_s(m_Z^2)}
\newc{\swsq}{\sin^2\theta_W}
\newc{\tw}{\tan\theta_W}
\newc{\cw}{\cos\theta_W}
\newc{\sw}{\sin\theta_W}
\newc{\BR}{\hbox{\rm BR}}
\newc{\zbb}{Z\to b\bar}
\newc{\Gb}{\Gamma (Z\to b\bar b)}
\newc{\Gh}{\Gamma (Z\to \hbox{\rm hadrons})}
\newc{\rbsm}{R_b^\hbox{\rm sm}}
\newc{\rbsusy}{R_b^\hbox{\rm susy}}
\newc{\drb}{\delta R_b}

\newc{\tbeta}{\tan\beta}
\newc{\uL}{{\tilde u_L}}
\newc{\uR}{{\tilde u_R}}
\newc{\cL}{{\tilde c_L}}
\newc{\cR}{{\tilde c_R}}
\newc{\tL}{{\tilde t_L}}
\newc{\tR}{{\tilde t_R}}
\newc{\dL}{{\tilde d_L}}
\newc{\dR}{{\tilde d_R}}
\newc{\sL}{{\tilde s_L}}
\newc{\sR}{{\tilde s_R}}
\newc{\bL}{{\tilde b_L}}
\newc{\bR}{{\tilde b_R}}
\newc{\eL}{{\tilde e_L}}
\newc{\eR}{{\tilde e_R}}
\newc{\mhp}{m_{H^\pm}}
\newc{\mhalf}{m_{1/2}}
\def\NPB#1#2#3{Nucl. Phys. {\bf B#1} (19#2) #3}
\def\PLB#1#2#3{Phys. Lett. {\bf B#1} (19#2) #3}

\def\PRD#1#2#3{Phys. Rev. {\bf D#1} (19#2) #3}
\def\PRL#1#2#3{Phys. Rev. Lett. {\bf#1} (19#2) #3}

\def\beq{\begin{equation}}
\def\eeq{\end{equation}}
\def\bea{\begin{eqnarray*}}
\def\eea{\end{eqnarray*}}
\def\slashchar#1{\setbox0=\hbox{$#1$}           % set a box for #1
   \dimen0=\wd0                                 % and get its size 
   \setbox1=\hbox{/} \dimen1=\wd1               % get size of /
   \ifdim\dimen0>\dimen1                        % #1 is bigger 
      \rlap{\hbox to \dimen0{\hfil/\hfil}}      % so center / in box 
      #1                                        % and print #1
   \else                                        % / is bigger 
      \rlap{\hbox to \dimen1{\hfil$#1$\hfil}}   % so center #1
      /                                         % and print /
   \fi}                                         %
\catcode`@=11
\long\def\@caption#1[#2]#3{\par\addcontentsline{\csname
  ext@#1\endcsname}{#1}{\protect\numberline{\csname
  the#1\endcsname}{\ignorespaces #2}}\begingroup
    \small
    \@parboxrestore
    \@makecaption{\csname fnum@#1\endcsname}{\ignorespaces #3}\par
  \endgroup}
\catcode`@=12

\begin{document}
\draft
 
\pagestyle{empty}

\preprint{
\noindent
%\begin{minipage}[t]{3in}
%\begin{flushleft}
%\today \\
%\end{flushleft}
%\end{minipage}
\hfill
\begin{minipage}[t]{3in}
\begin{flushright}
%DRAFT \\
%Not for distribution \\
SLAC--PUB--7451 \\
hep-ph/9706085 \\
May 1997
\end{flushright}
\end{minipage}
}

\title{Multi-GeV photons from electron--dark matter \\
scattering near Active Galactic Nuclei}

\author{
Elliott D.~Bloom, James D.~Wells
\thanks{Work supported by the Department of Energy under Contract
DE--AC03--76SF00515.}
} 

\address{
Stanford Linear Accelerator Center,
Stanford University, Stanford, California 94309
}

%\date{\today}
\maketitle

\begin{abstract}

Active Galactic Nuclei (AGN) may emit highly collimated and intense jets of 
relativistic electrons which upscatter ambient photons.  
These electrons can also scatter off the
cold dark matter halo of the galaxy to produce high energy photons
which have a more isotropic signature than the up-scattered photons from
QED processes.  
We propose to look for these
high energy photons coming from AGN as a method to detect dark matter.
As a primary example
we work out the expected signal from electrons scattering
off the lightest supersymmetric partner into a photon plus selectron.  
Using the optimistic side of astrophysical uncertainties, we still
find the signal from M87 or Centaurus A, two close-by AGN,
smaller than the sensitivities expected of
the currently proposed photon detectors.  However, long running
photon detectors and future detectors of higher sensitivity
might be able to distinguish
a signal from AGN sources.  In order to have confidence that new physics
sources are discernible, we also emphasize the importance of multi-wavelength
studies of AGN with varying jet axis orientation to the earth.

\end{abstract}
%\pacs{PACS numbers}
%\newpage
\pagestyle{plain}
\narrowtext

\setcounter{footnote}{0}

\baselineskip=33pt

Astrophysics observations and cosmological arguments point to the
necessity for a substantial amount of cold dark matter in the
universe~\cite{wimps}.  It is even possible that
the invisible cold dark matter constitutes
nearly critical density ($\rho_c = 3H^2/8\pi G$) in the standard
$\Omega =1$ big-bang cosmology.  Dark matter detection to date
has only been accomplished by witnessing gravitational effects
(rotation curves of galaxies, Hubble flow distortions, gravitational
lensing, etc.).  
Since the dark matter is probably
not electrically charged or color charged~\cite{charged},
its most likely non-gravitational interactions are due to the weak force.
Hence, the hypothesis of stable weakly interacting massive particles
(WIMPs) has been considered~\cite{wimps}.  
We will use the bino (superpartner of the hypercharge gauge boson)
as the lightest supersymmetric partner (LSP) in our example
scattering process.

Experimental effort is now underway on many fronts~\cite{susydm} to see
dark matter via its weak interactions.  Cryogenic table top experiments which
measure nuclear recoil of ambient LSPs interacting with Silicon, Germanium,
and other elements are progressing.  One can also search
for neutrino yields from
the annihilation of captured LSPs in the core of the sun or earth.  And
searches are underway to look for
positrons, electrons, protons and antiprotons from
the annihilations of LSPs in the galactic halo.

Here, we propose another idea to search for LSPs: photons originating in
the final state of electron-LSP scattering near the center of Active
Galactic Nuclei (AGN).  The principle property of an AGN of the
Blazar type~\cite{blazars} which makes
this idea potentially feasible is the high energy 
and high flux electron
beams which are thought to be emanating from their 
centers~\cite{dermer,blazars}.  
The high energy jets are thought to arise from a highly relativistic shock
wave traveling along the perpendicular axis of an accretion disk
associated with a massive black hole at the center ($10^6-10^9~M_{\odot}$).
Shock waves generated by the central engine accelerate
electrons and positrons which in turn
produce high energy photons mainly through inverse Compton scattering
of the~UV radiation from a thermal distribution of photons originating
external to the electron/positron
jet in the accretion disk~\cite{dermer}.  Other processes
that could be important include inverse Compton
scattering of synchrotron radiation, and pair annihilation
of the electrons and positrons into high energy photons.  We will loosely
call all these processes ``up-scattering'' of high energy
electrons and positrons into high energy photons.

The lab frame, which is the accretion disk frame, or equivalently 
our observation
frame on earth (for close-by AGN),
has a distribution of electrons which scales roughly
as $E^{-2}$, and is highly collimated with a beaming factor (solid angle
fraction) of 
about $10^{-3}$~\cite{dermer}.  
Since the upscattered radiation can be measured, the coefficient
out in front of the assumed 
$E^{-2}$ can be measured to obtain an estimate of the
electron/positron flux ejected from the AGN.  

The total power of the AGN in photon radiation is approximately
$\Delta F_\gamma = 10^{47}f - 10^{49}f~\mbox{ergs}/s$ in the energy range of
$100\mev <E_\gamma < 5\gev$~\cite{fichtel}, where $f$ is the 
beaming factor (fraction
of solid angle) of the electron beam. We shall take $f\simeq 10^{-3}$ in
accordance with~\cite{dermer}.
Since the hadronic component of the AGN jet is probably
less important~\cite{blazars} 
to the photon radiation than the electron/positron constituents,
we will make the approximation that all the photon radiation power is
converted from the electron jet (often we use ``electron'' to mean
both electrons and positrons).  
This enables us to estimate the intensity
and energy distribution of the electron beam.  We find the approximate
value by using~\cite{dermer} 
the same energy profile for the electrons as the photons:
\begin{equation}
\label{eflux}
\frac{df}{dE_e}=n_0 \left( \frac{m_e}{E_e}\right)^2 .
\end{equation}
We can estimate the value of $n_0$ to be
\begin{equation}
n_0\simeq \frac{\Delta F_\gamma}{m^2_e}\frac{1}{\log 5\gev /100\mev}
 \simeq 10^{55} /\gev /s.
\end{equation}
Given this normalization, and given this distribution
between $1\mev$ to $10\tev$, the total amount of energy ejected
from the AGN is about one solar mass per year.  It might be that much
more energy is ejected from the AGN, and that all the electron beam is not
transmitted into photons, in which case the above estimate for $n_0$ would
be too low.  However, it should also be kept in mind that there are other
models of high energy photon jets from AGNs which do not require such intense
beams of electrons~\cite{protons}.

If all the electron power went into
merely up-scattering photons through simple QED processes, then there would
be no electrons left to interact with dark matter.  
Below it will become apparent that the penetration depth of the electrons
is not crucial unless it is less than a few parsecs from the central
engine, since almost all
the electrons have to interact with the LSPs within about ten parsecs
of the accretion disk in order for a signal to be viable at all.
Together with the observation that a relatively small fraction of the 
electrons do scatter off the LSPs, we are free to use Eq.~\ref{eflux} 
as our particle
beam source which interacts with the dark matter.  Although there are some
reasons to believe that the electrons could upscatter the photons far away
from the central engine~\cite{sikora}, the reader should keep in mind that 
a detailed model description of the electron/photon beam will involve
a distance dependent attenuation factor for the electron beam.
Current models are not precise enough to calculate the attenuation
so we have assumed optimistically that it is negligible.

The dark matter target distribution must also be 
modeled~\cite{other cdm,core cdm}.  Several differing
proposals are in the literature.  Here, we choose to model
the dark matter distribution after Ref.~\cite{berezinsky} which postulates
that $\rho\sim r^{-1.8}$.  We choose this model over the more traditional
distributions of $\rho \sim 1/(r^2+r_c^2)$, where $r_c$ is a hard-core density
radius which smoothes $\rho$ near $r=0$, for 
two reasons:  (1) The lack of a hard-core density radius in this model
is becoming more and more observed in numerical simulations~\cite{core cdm}, 
and (2) large
enhancements of $\rho$ near $r\sim 0$ are needed for our proposed signal 
to be interesting.  That is, without
a cusp in the dark matter distribution~\cite{ipser}, 
or equivalently, without $r_c\lsim$ few
parsecs, our resulting signal flux of photons 
would be too low to be detected.

To be explicit, we use the dark matter distribution 
\begin{equation}
\rho (r) = \rho_a \left( \frac{a}{r}\right)^{1.8}
\end{equation}
where $a$ is some arbitrary distance from the center of the galaxy and
$\rho_a$ is the local density at that distance.  For numerical purposes
we will use $a=8$ kpc and $\rho_a = 0.3\gev /\mbox{cm}^3$, in accord with
the density profile of our galaxy.  It is important to estimate the minimum
radius size, $r_{min}$, at which this density function breaks down.  The
different physics sources that could disrupt this density profile are
given in Ref.~\cite{berezinsky}, and it was determined that a 
central black hole would yield the largest $r_{min}$ of $\sim 1\, pc$
(for $M\sim 10^8 M_\odot$).
The assumption of a black hole at the center of the galaxy is quite
applicable to our AGN study, and we adopt the black-hole hypothesis 
as the origin of $r_{min}$.  The capture radius of 
dark matter is then between about 0.01~pc to 10~pc 
depending on black hole mass,
dark matter density, etc.  For numerical purposes we will use both
1~pc and 1~kpc.  

\begin{figure}
\psfig{file=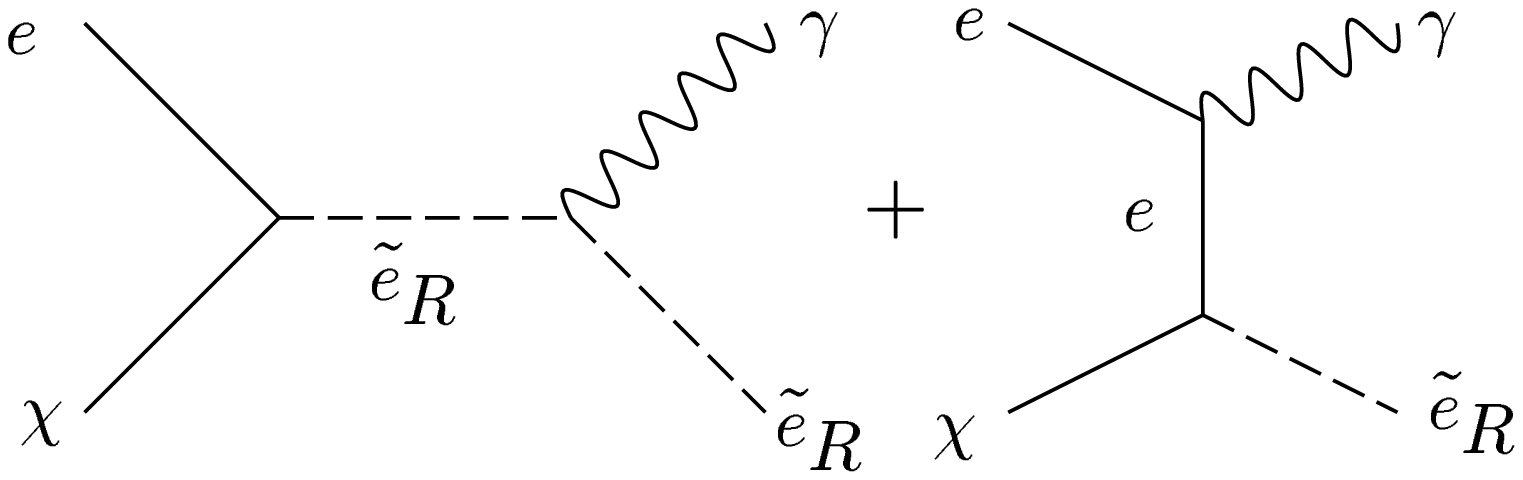}
\caption{Feynman diagrams contributing to $e^\pm\chi\to \gamma\tilde e_R$
scattering.}
\label{ebino}
\end{figure}

We now have modeled both the target (dark matter distribution) and the
source (high energy electron beam), and so we can estimate the flux
of photons originating from $e^\pm \chi \to \tilde e_R^\pm \gamma$ scattering.
Our notation identifies the bino with $\chi$ and the right-handed
selectron (superpartner of the right-handed electron) as $\tilde e_R$.
Feynman diagrams for the process are given in Fig.~\ref{ebino}.

Before setting up the differential flux, it is instructive to summarize
some basic relativistic kinematics and notation.  For close-by AGN,
the lab frame of
the electron/LSP collision is the observer frame. 
Thus $E_e$ is the incident electron energy, making the center of
mass energy equal to 
\begin{equation}
s= m_\chi (m_\chi +2E_e),
\end{equation}
and the resulting photon energy in the lab frame is
\begin{equation}
E_\gamma = \frac{s-m^2_{\tilde e_R}}{2m_\chi}.
\end{equation}

In the c.m. frame (primed quantities) the electron and photon
energies are
\begin{equation}
E_e' =  \frac{s-m^2_\chi}{2\sqrt{s}} ~~~
E_\gamma'  =  \frac{s-m^2_{\tilde e_R}}{2\sqrt{s}}.
\end{equation}
Relativistic transformations of four vectors in the lab frame to the
bino/electron c.m. frame are carried out by 
\begin{equation}
\beta = \frac{E_e}{E_e+m_\chi}.
\end{equation}
The angle $x=\cos\theta$ with respect to $\hat z$ (jet axis)
in the lab frame is related to the angle
$x'=\cos\theta'$ in the c.m. according to
\begin{equation}
\label{xtrans}
x = \frac{\beta +x'}{1+\beta x'}.
\end{equation}

In the c.m. frame the differential cross-section for
$e\chi\to \tilde e\gamma$ scattering in the limit
that $\sqrt{s}\gg m_\chi,m_{\tilde e_R}$ simplifies to
\begin{equation}
\frac{d\sigma}{d\Omega'} \simeq \frac{\alpha^2}{2\cos^2\theta_W}\frac{1}{s}
\left( \frac{1+\cos\theta'}{1-\cos\theta'} \right),
\end{equation}
where $\alpha=1/137$ is the QED coupling constant, and $\theta_W$ is
the weak mixing angle.
(Due to finite electron mass and selectron width there is no singularity
in the cross-section as $\theta'\to 0$.)
We keep the masses of $\chi$ and $\tilde e_R$ in our numerical work.
To cast this in the lab frame we use Eq.~\ref{xtrans} to substitute
$x$ for $x'$ and to
calculate $dx'/dx$ as a function
of $x$.  

The total differential $e^\pm \chi$ scattering flux 
observed at the earth can be parameterized as
\begin{equation}
\label{almost}
\frac{dF}{dE_e} = \frac{\mbox{\#}~\mbox{events}}{cm^2 sec\gev} =
\frac{1}{d_{AGN}^2}\left[ \frac{d\sigma}{d\Omega}\right]_{x_0} 
  \frac{d{\cal L}}{dE_e},
\end{equation}
where $x_0=\cos\theta_0$ is the fixed angle between the jet axis and the
axis which points from the AGN to the earth.  Also,
$d_{AGN}$ is the distance to the AGN and ${d{\cal L}}/{dE_e}$ is
the effective luminosity given by
\begin{equation}
\frac{d{\cal L}}{dE_e}=\frac{df}{dE_e}\frac{\langle \rho_\chi l_\chi\rangle}
 {m_\chi}.
\end{equation}
The factor $\langle \rho_\chi l_\chi\rangle$ is the average of the
$r$ dependent dark matter
density function times the effective length that the electrons
pass through, and is defined as
\begin{equation}
\langle \rho_\chi l_\chi \rangle = \int_{r_{min}}^{r_{max}} dr \rho(r) \simeq 
 r_{min} \rho_a \left( \frac{a}{r_{min}}\right)^{1.8}\simeq 
 r_{min}\rho(r_{min})
\end{equation}
where the latter approximation is for $r_{max}\gg r_{min}$.

There is a one-to-one correspondence between the electron
energy $E_e$ and the photon energy $E_\gamma$ given by
\begin{equation}
\label{eeeg}
E_e = \frac{m^2_{\tilde e}-m^2_\chi+2m_\chi E_\gamma}{2m_\chi}.
\end{equation}
Since $dE_e/dE_\gamma = 1$ the differential
photon flux is given simply by substituting $E_e$ of
Eq.~\ref{eeeg} into Eq.~\ref{almost}.  The integral photon flux is then
\begin{equation}
F(E_\gamma)= \mbox{\#}~\mbox{photons}\, \mbox{cm}^{-2}\, \mbox{sec}^{-1}
 = \int_{E_\gamma}^{\infty} dE_\gamma'' \frac{dF}{dE_\gamma''}.
\end{equation}

Figure~2 shows the results for the AGN M87 which is approximately 12~Mpc
away, and has a jet axis oriented at $42\pm 5$ degrees to 
us~\cite{elliott,elliottref25}.  
We also show the result for Centaurus A, which is $\sim 2.5$ Mpc away and
is oriented at $\sim 68$ degrees~\cite{jones}.
The supersymmetric
model used to construct Fig.~2 
is a pure bino LSP with mass 60~GeV, and a right-handed
selectron with mass 100~GeV.  The upper solid line is with $r_{min}=1\, pc$
and the bottom solid line is with $r_{min}=1\, kpc$.  
The top dashed line is the
the $5\sigma$ sensitivity of the proposed space based gamma ray
telescope GLAST~\cite{GLAST}, 
and the lower right line
is the expected sensitivity of the proposed ground based gamma
ray telescope array VERITAS~\cite{VERITAS}.  For GLAST, the sensitivity
corresponds to one year scanning mode in which the instrument
is always pointed outward
from the earth (approximately 30\% duty cycle).  For VERITAS, the
sensitivity is for one week observing time on the source.
As we can see, the signal is
lower than the present sensitivities of these proposed photon detectors.  

The AGN beam as modeled in~\cite{dermer} has a non-zero gamma ray flux
at large angles from the jet axis.  For Centaurus A the flux
at 68 degrees off jet axis (i.e., earth direction is 68 degrees 
from jet axis) is
expected to be
negligible; the electron/positron beam is highly collimated, dramatically
decreasing
the overall flux normalization and
especially
decreasing the gamma ray energy cut-off from ordinary QED processes
with higher observing angle.
However, for M87 whose jet axis is only $42\pm 5$ degrees
with respect to the earth direction, the intrinsic background is substantial
and swamps the expected flux from 
electron/bino scattering up to approximately $150\gev$.  
In the figure
we have plotted this background estimate using the same photon
beam flux used earlier in deriving the dark matter signal,
and we have also assumed a 45 degree jet axis orientation for
easier comparison with ref.~\cite{dermer}.
Measuring gamma ray
fluxes from intermediate angle ($10^\circ \lsim \theta \lsim 50^\circ$
off jet axis)
AGN such as M87 are still useful because they
can increase our understanding of AGN jets, and possibly the model
of ref.~\cite{dermer} 
in particular.  Multi-wavelength measurements of the photon flux from a number
of AGN with different jet axis orientations is highly desirable to more
fully characterize the AGN beams including determination of the angle
of the jet
axis to the earth using radio measurements.  
Such measurements will allow deeper
understanding of AGN beams and possibly allow us to have confidence that
new physics is discernible.

Although we typically chose parameters such as $r_{min}$ and $n_0$ which
gave maximal contribution to the signal, there are many astrophysical
uncertainties involved in the calculation that could make the signal perhaps
larger 
or smaller than we have estimated here. Summing over all 
large angle Blazar sources
will of course help increase the signal over the diffuse background.  
Such techniques have already
been shown to have a significant impact on sensitivity in the
X-ray region~\cite{refregier}.

We have focused on the case of a pure bino dark matter 
candidate.  Certainly other particles are around, even within
supersymmetry, that yield good dark matter candidates.  These candidates
could of course have much different scattering amplitudes than the bino.
For example, the higgsino dark matter candidates discussed in~\cite{higgsino}
would not allow a direct coupling of the higgsino to an electron and selectron.
One useful final state is a chargino and a neutrino produced
from $t$ channel $W$ exchange. The decay chain from the chargino then can
include a photon from 
$\chi^0_2\to \chi^0_1\gamma$.  Such decays could lead to structure
in the photon energy spectrum, providing an additional signature.
In any event, the final cross-section will
have electroweak strength and should not be significantly different
than the bino case studied above.

We have found that our estimate of the signal for electron-bino
scattering into photons from a single AGN source is too low
to be discerned by currently proposed experiments, but future
detectors of much higher sensitivity might be able to see a signal.
The limit on sensitivity to the photon signal of the most sensitive
of the currently proposed experiments -- ground based Cherenkov 
telescope arrays -- is largely due to the difficulty of subtracting
pseudo-backgrounds primarily from electron showers in the atmosphere.
Therefore, if a signal does exist it might be possible to reach the
sensitivities required by the continuing progress to eliminate these
backgrounds, and dramatically extend the observing time on candidate
sources.  Obviously this presents great challenges.

%%%%%%%%%%%%%%%%%%%%%%%%%%%%%%%%%%%%%%%%%%%%%%%%%%%%%%%%%%%%%%%%
\acknowledgements

We would like to thank M.~Crone,
C.~Dermer, and M.~Kamionkowski for useful discussions.  We also thank
C.~Dermer for providing us an estimate to the electron flux 
energy distribution
in the AGN jets. We thank the Aspen Center for Physics where the authors
originated this work last summer.

%%%%%%%%%%%%%%%%%%%%%%%%%%%%%%%%%%%%%%%%%%%%%%%%%%%%%%%%%%%%%%%%

%%%%%%%%%%%%%%%%%%%%%%%%%%%%%%%%%%%%%%%%%%%%%%%%%%%%%%%%%%%%%%%%

\input psfig
\noindent
%\vfill\eject

\noindent
\begin{figure}
\psfig{file=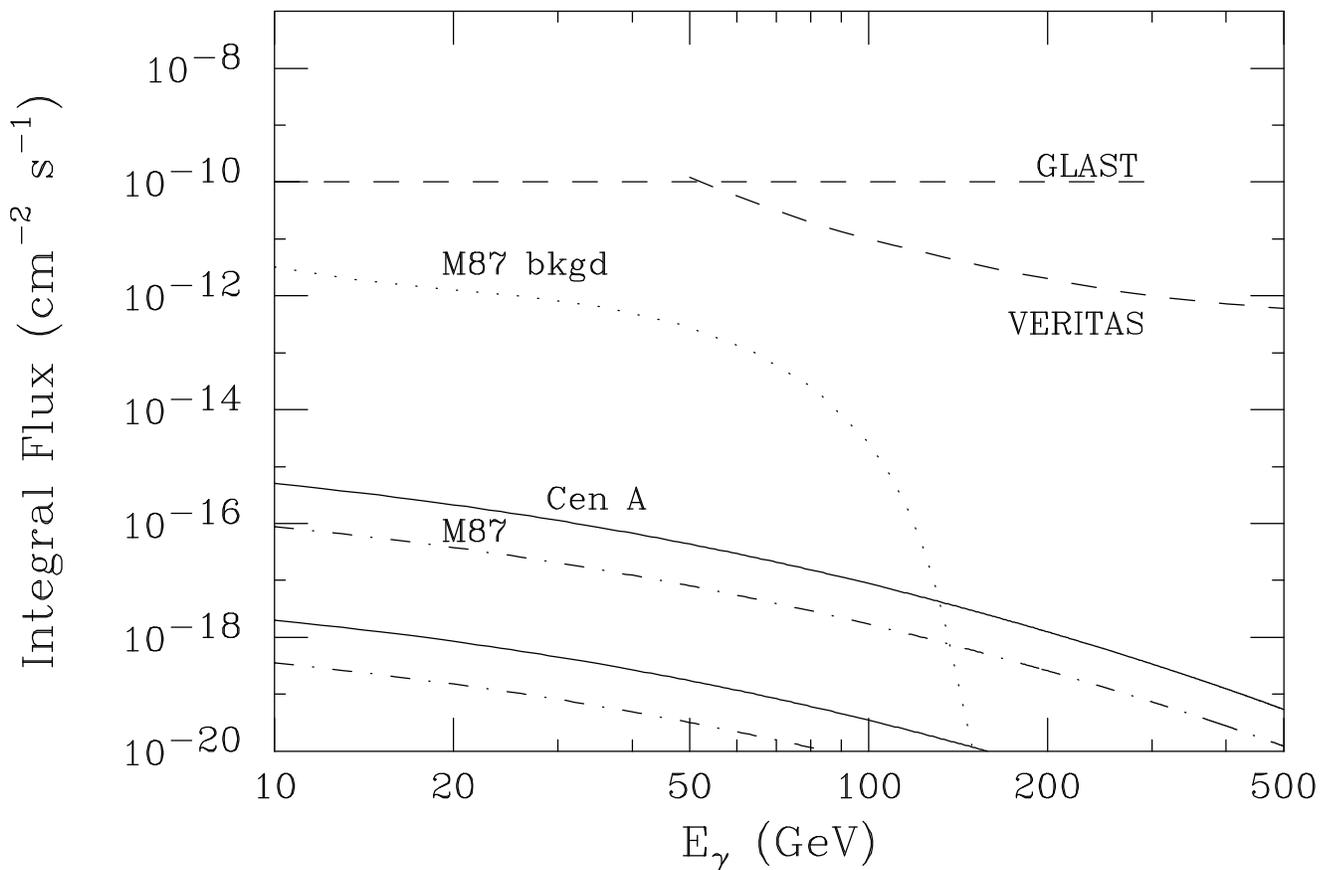}
\caption{Integral flux of photon signal from the Centaurus A and M87 Active
Galactic Nuclei.  The top solid (dash-dot) line corresponds to $r_{min}=1$~pc
in the dark matter density profile for Centaurus A (M87), and the bottom
line corresponds to $r_{min}=1$~kpc.  The dashed lines represent the 
sensitivity of the different photon detectors GLAST (one year scanning
mode) and VERITAS (one week on source).
The dotted line represents an estimate of the background flux from
the M87 jet.  Gamma ray jet background from Centaurus A is expected to
be negligible.   See the text for further discussion.}
\label{fig:1}
\end{figure}
 
\end{document}